# THE PULSE OF THE CITY THROUGH TWITTER: RELATIONSHIPS BETWEEN LAND USE AND SPATIOTEMPORAL DEMOGRAPHICS

Juan Carlos García-Palomares, María Henar Salas-Olmedo, Borja Moya-Gomez,

Ana Condeço-Melhorado and Javier Gutierrrez

*Departamento de Geografía Humana, Universidad Complutense de Madrid, Spain*

**Abstract.** Social network data offer interesting opportunities in urban studies. In this study, we used Twitter data to analyse city dynamics over the course of the day. Users of this social network were grouped according to city zone and time slot in order to analyse the daily dynamics of the city and the relationship between this and land use. First, daytime activity in each zone was compared with activity at night in order to determine which zones showed increased activity in each of the time slots. Then, typical Twitter activity profiles were obtained based on the predominant land use in each zone, indicating how land uses linked to activities were activated during the day, but at different rates depending on the type of land use. Lastly, a multiple regression analysis was performed to determine the influence of the different land uses on each of the major time slots (morning, afternoon, evening and night) through their changing coefficients. Activity tended to decrease throughout the day for most land uses (e.g. offices, education, health and transport), but remained constant in parks and increased in retail and residential zones. Our results show that social network data can be used to improve our understanding of the link between land use and urban dynamics.

## 1. INTRODUCTION

Internet users are no longer mere passive recipients of information, but have become producers of vast amounts of data, particularly through social networks. Many social media apps (Twitter, Foursquare and Facebook are three common examples) have geo-location features that (optionally) includes the ability to attach locational information in the form of coordinates provided by the units GPS or place names provided by the user, thereby enabling individual users to leave a digital 'geographic footprint' of their movement when posting a message (Blanford et al., 2015). These digital data shadows are intimately intermingled with offline, material geographies of everyday life (Shelton et al, 2015). High spatial and temporal social media data reveals activity patterns information moving beyond the nighttime residential geographies of conventional geodemographic data sources (Longley et al., 2015).

Twitter has become an important sensor of the interactions between individuals and their environment (Frias-Martinez, V., & Frias-Martinez, E. (2014). Twitter data contains precise spatial and temporal information which can be used to perform spatiotemporal demographic analyses. It has been observed that residential zones on the periphery of cities generate more tweets in the evening, when people have returned to their homes, whereas areas of activity in the city centre are especially active during the day, when people visit them to undertake activities such as work or shopping (Ciuccarelli et al., 2014). The spatiotemporal pattern of tweets posted from different zones in the city indicates the existence of the inherent linkage between the human urban activity pattern and the underlying land use structure (Zhan et al., 2014).

The aim of this study was to use information from the social network Twitter in order to analyse city dynamics throughout the day and the relationship between this and land use. We estimated the number of unique active users in each zone of the city of Madrid throughout the day, as a proxy for the changing location of the population. Then, we analysed Twitter activity (active users) throughout the day according to land use, yielding typical activity profiles for each land use. Lastly, we calculated multiple regression models (OLS) for each time slot to explain Twitter activity based on square metres of each land use in each zone of the city. The underlying hypothesis was that if each land use possessed a characteristic Twitter activity profile, then the surface area of each land use should satisfactorily explain the distribution of tweeters in the city in each time slot. An analysis of the coefficients of the independent variables should indicate the changing influence of the different land uses according to time slots in agreement with the typical activity profiles previously obtained.

In contrast to previous studies that have used clustering algorithms for city zones according to the Twitter activity taking place and then relate the classification obtained to the land use, in our study we analysed the specific time profiles of each land use based on a prior classification of land use for urban planning and then used explanatory models that linked land use and Twitter activity. Another distinguishing feature of our study is that the unit of analysis was the Twitter user rather than the geotagged tweet. We focused not on the spatiotemporal distribution of tweets but on that of Twitter users, as a proxy for the changing location of the population throughout the day. This approach mitigates the bias derived from different intensities of Twitter use, which gives more weight to users who tweet more often, and above all, it provides useful information for the provision of public sector services and private sector business activities.

The remainder of this paper is structured as follows. Section 2 presents a literature review on the use of Twitter data in urban studies, especially in relation to land uses. Section 3 reports the methodology and data. Section 4 contains an analysis of the results and section 5 presents the main conclusions.

## 2. HUMAN ACTIVITIES AND LAND USES: SPATIOTEMPORAL PATTERNS OF DIGITAL FOOTPRINTS IN THE CITY

Social network users generate a vast amount of data. Worldwide, 500 million Twitter messages are posted every day, there are 7,000 million check-ins on Foursquare, and more than 80 million photos are uploaded on Instagram. Most studies based on social network data have used Twitter (Murthy, 2013), due in large part to the fact that the data (tweets) can be freely downloaded by connecting to the Twitter Streaming API. This realtime, public, and cost-free data stream covers only around 1–2 percent of the whole stream of tweets (Andrienko et al., 2013). Some of the tweets are geotagged. These facilitate profiling of usage across space as well as time and have been found to be a useful tool for urban research (Lansley and Longley, 2016).

Within a city from the centre to the periphery, the densities decrease in general, so tweet densities would be a good surrogate of population densities (Jiang et al., 2016). Maps of the spatial distribution of tweets can be enriched with user IDs, tweet content and language. Thus, Longley et al. (2015) inferred the gender, age and ethnicity of tweeters from the user ID (username field) and analysed the spatial distribution of the tweets posted by each of the groups identified. Lansley and Longley (2016) and Andrienko et al. (2013) focused on tweet content in order to examine frequently tweeted words and their spatiotemporal patterns. Meanwhile, Mocanue et al. (2013) used the language of tweets to identify linguistically specific urban communities. It is also possible to analyse the degree of social mixing in the use of space, tracking the movement of demographic groups within cities (Netto et al., 2015; Shelton et al., 2015). In contrast to the information provided by official sources, which offer data on place of residence, these studies on multiculturalism and social mixing analyse the use of space throughout the day.

Through spatiotemporal monitoring of the tweets posted by each user, it is possible to deduce population mobility patterns (Wu et al., 2014; Blanford et al., 2015; Longley and Adnan, 2016), identify demographic groups based on user names (Luo et al., 2016) and obtain origin-destination matrices (Gao et al., 2014). The reliability of Twitter data in mobility studies has been validated by Lenormand et al. (2014), who compared Twitter data with mobile phone records and official data (census) and concluded that the three sources offer comparable results.

The spatial distribution of tweets in the inner city changes over the course of the day, reflecting changes in the location of the population. Areas of the city with similar time profiles can be grouped using clustering algorithms in order to identify clusters of common tweeting activity. Thus, Frías-Martínez et al. (2012) and Frías-Martínez and Frías-Martínez (2014) identified four clusters with specific tweeting activity signatures that basically corresponded to the following types of land use: business, leisure/weekend, nightlife and residential. Using a similar methodology, Zhan et al. (2014) inferred four broad categories of land use based on a spatiotemporal analysis of Twitter data: residential, retail, open space/recreation and transportation/utility. A similar approach has been adopted in other studies (e.g., Pei et al., 2014; Ríos and Muñoz, 2017), in which clustering algorithms have been used to group city zones with similar profiles according to mobile phone activity, or Chen et al. (2017) in China using the social media "*Tencent*". These methodologies based on clustering the tweeting activity profiles of different areas in the city could be used as an alternative to satellite imagery, to infer urban land uses based on social network data. However, the land use categories obtained are very generic, and their usefulness for urban planning is therefore limited.

## 3. DATA AND METHODOLOGY

### 3.1 Downloading and processing data

One of the most widespread social networks is undoubtedly Twitter, a platform for posting messages with a maximum of 140 characters, known as tweets. The service has more than 270 million active users around the world. Roughly 80% of active Twitter users access the service via a mobile telephone (Lansley and Longley, 2016). Since 2010, Twitter provided users with the ability to include their location either by attaching coordinates or a place name while tweeting, therefore making it possible to locate tweets geographically both in space and over time (Blanford et al., 2015). Tweets are automatically geotagged by the GPS of mobile devices, provided that the user has enabled this function. Geotagged Tweets account for approximately just 1% of all the messages that are sent using the Twitter service (Blanford et al., 2015).

Twitter data was downloaded and pre-processed as follows:

*a) Data download*

The data used for this study was downloaded via the Twitter Streaming API over two consecutive years (from January 2012 to December 2013). Only geotagged tweets were downloaded, selecting those that covered the municipality of Madrid. Besides the coordinates, each tweet also carries information about the user ID, the date and time the tweet was posted, the device language setting, device type and the text of the message. We downloaded over 6.8 billion tweets in the city of Madrid. The data were loaded into a GIS (ArcGIS 10.4), creating a layer of points with the coordinates *x y* of the position from which each of them was posted. Subsequently, we selected tweets that had been posted on typical workdays (Tuesday, Wednesday and Thursday), obtaining a total of 3.07 million tweets.

*b) Spatial and temporal aggregation of data to obtain the number of unique active users*

The same user often posts several tweets from the same location at the same time. The number of such tweets can be extremely high with some users, leading to an overestimation of the presence of this type of user at these locations and times. It is therefore necessary to analyse unique users rather than tweets. To this end, the tweets were aggregated spatially and temporally (every quarter of an hour) depending on the user ID, to obtain the presence of unique active users in each spatial unit rather than the number of tweets posted. This approach eliminates compulsive users who remain in the same place.

Spatial aggregation of Twitter data was based on the spatial zones established by the Madrid transport authority. This zoning was chosen because although most transport zones present some degree of mixed land use, in general they form homogeneous spatial units from the point of view of land use and building typology. In addition, these zones are sufficiently large to register a significant number of tweets. The possibility of using cells or hexagons (Shelton et al., 2015) was ruled out. Cells undoubtedly have the advantage of mitigating the problem of modified spatial units (Openshaw, 1984), since they form spatial units of the same size and shape; however, such units are not homogeneous from the point of view of land use, and the aim of our research was to relate the number of Twitter users to land use. The temporal sequence obtained of the presence of unique active users in each transport zone every quarter of an hour can be viewed in the attached video, which shows how hot spots shifted from the centre, in the morning and the afternoon, towards the periphery, at night.

To relate the distribution of unique active Twitter users to land use, each transport zone was characterised in terms of percentage of surface area occupied by each land use, based on land registry records. Initially, three main types of land use were distinguished: residential (when more than 66.6% of built surface in the area was residential), activity (when more than the 66.6% was non-residential, e.g. offices, industry, retail or education) and mixed residential (all other cases). In addition, the areas of activity were classified into 9 types: offices, industry, retail, health, education, culture, large transport terminals, parks and others. Figure 1 shows the predominant land use in each of the zones.

**Figure 1: Predominant land uses according to transport zone**

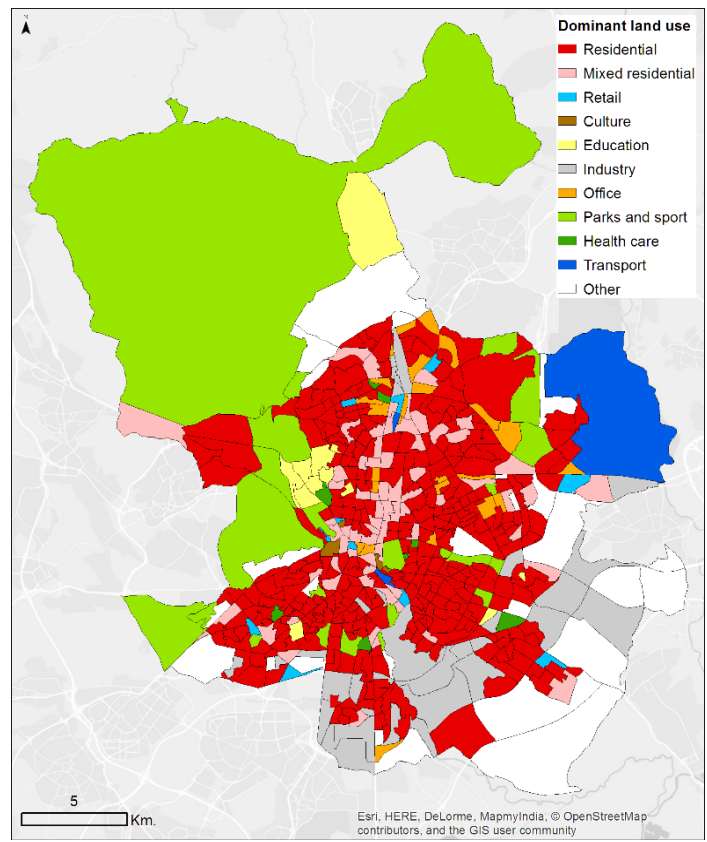

*c) Normalisation of data and calculation of the number of users per transport zone and time of day*

One of biases presented by Twitter when using temporal information is the different use made of this social network over the course of the day. Twitter is used more in the evening than in the morning or at mid-day (Figure 2). The highest peak in users occurred at ten pm (over 7%), whereas this figure fell to less than 4% at 8-9 am.

**Figure 2. Distribution of the number of Twitter users in Madrid according to time slot**

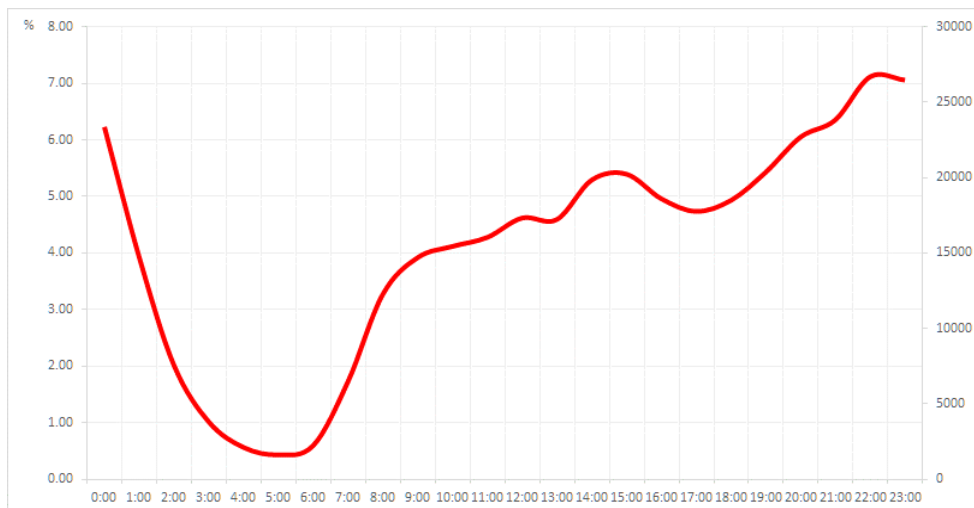

To eliminate this bias, daily distribution data were grouped by time slot and the distributions were normalised, so that the total number of unique active users per time slot was equalised to 100,000, using the following formula:

$$T_{zhn} = \frac{T_{zh}}{T_h} * 100000,$$

where $T_{zhn}$ is the normalised number of unique active users in zone *z* at time slot *h*; $T_{zh}$ is the number of unique active users in zone *z* at time slot *h* and $T_h$ is the total number of active tweeters at time slot *h*.

Subsequently, the data were aggregated into four time slots to facilitate analysis: a) morning (07:00 to 12:59), b) afternoon (13:00 to 17:59), c) evening (18:00 to 22:00) and d) night (22:00 to 23:59)[1].

### 3.2 Analysis of spatial distribution patterns

Initially, the spatial distribution patterns were analysed by mapping unique active users according to large time slots. Descriptive statistics were obtained for these distributions. In this case, since the data were normalised, the mean distributions were similar in all cases, but their standard deviations showed the degree of concentration or dispersion of Twitter users. High standard deviation values are associated with more concentrated spatial patterns (users are concentrated more in some zones and less in others). In contrast, low standard deviation values indicate user dispersion (more uniform distribution throughout the zones).

To compare the distributions according to time slot, we conducted a bivariate Ordinary Least Squares (OLS) analysis. ArcGIS10.4 software was used to obtain both the coefficient of determination and the spatial distribution of the residuals. The coefficient of determination indicates the degree of relationship between two variables (the overlap between data distributions), while an analysis of normalised residual plots identifies differences between the distributions. Although we analysed all the relationships between the four major time slots, we focused primarily on comparing night-time distribution (as a baseline reference) with each of the other time slots.

### 3.3 Temporal profiles according to land use

Temporal profiles according to land use were calculated using the normalised distributions of unique active users according to zone every quarter of an hour and the predominant land use in each zone. The total number of unique active users every quarter of an hour in a zone was assigned to the predominant land use, and then the total number of users according to land use was summed for each quarter of an hour. This approach yielded the temporal distribution of unique active users for each land use.

---

[1] The assumption that the normalised number of unique active users reflected the location of the population should be contrasted with the ground truth. Official data record the location of the population according to place of residence (night-time residential geographies), but do not provide information on the location of the population over the course of the day. Therefore, a comparison with the ground truth can only be performed with night-time data (Lin and Cromley, 2015). As in previous studies (e.g. Luo et al., 2016; Salas-Olmedo and Rojas, 2017), we defined "home" as the place most frequently visited by a user at night-time. Thus, we included users who tweeted between 22:00 and 24:00 from residential buildings. In order to verify if the number of the detected residents out of the Twitter users in each transport zone adequately reflected the distribution of the population according to official data (Register of Inhabitants 2013, Spanish National Statistics Institute), we calculated the correlation coefficient between the two variables, obtaining an r2 value of 0.46, indicating that the number of resident tweeters is a good proxy for the distribution of the population.

### 3.4 Ordinary Least Squares (OLS) models

Lastly, the relationship between Twitter user distribution and land use was analysed using four multiple regression (OLS) models. The dependent variables were the normalised distributions of unique active users according to zone in each of the major time slots (Morning, Afternoon, Evening and Night). The explanatory variables were square metres of each type of land use in each of the zones in the city. Given the marked centre-periphery drop in the intensity of Twitter use (see subsection 4.1), distance to the city centre was included in the models as a control variable.

The OLS models were calculated in two steps. In the first step, we included all the independent variables considered initially (land uses and distance to the centre). In the second step, we eliminated non-significant variables and calculated the models again. The results shown in Table 4 correspond to the second step, and the variables that proved non-significant in the first step and were discarded have been left blank.

## 4. RESULTS

### 4.1 Distribution of tweeters according to time slot

Twitter activity in the different city zones varied substantially according to time slot. For the purpose of conducting comparisons, night-time was taken as the reference scenario (Figure 3). The differences in distribution according to time slot can be analysed using bivariate correlations (Table 1). As is to be expected in a place with a wide variety of land uses such as Madrid, the coefficients of determination between the time distributions were relatively high in all cases, especially between successive time slots in the day. Taking the night as the reference, the greatest differences (lowest correlations) were observed between night and morning (homes vs activities). In contrast, the correlations were highest between night and afternoon, and especially between night and evening (when many people have returned home after a day's work).

The residual plots of the correlations between night and the rest of the time slots show the different behaviour of residential spaces and areas of activity throughout the day (Figure 4). Comparing night and morning, areas of activity were more active in the latter slot (positive residuals in red), especially in office and mixed zones in the centre (with a high presence of offices) and in large complexes (e.g. university campuses and hospitals) and transport terminals (airport and railway stations). In contrast, residential areas presented negative residuals (in blue). Comparing night and evening, many areas of activity ceased activity while leisure centres and shopping areas became more active.

**Table 1. Relationships in the distribution of users according to time slot (r2)**

|  | Morning: 08:00 to 14:00 | Afternoon: 14:00 to 19:00 | Evening: 19:00 to 22:00 | Night: 22:00 to 24.00 |
|---|---|---|---|---|
| Morning: 08:00 to 14:00 | 1 | | | |
| Afternoon: 14:00 to 19:00 | 0.89 | 1 | | |
| Evening: 19:00 to 22:00 | 0.71 | 0.89 | 1 | |
| **Night: 22:00 to 24:00** | **0.49** | **0.72** | **0.81** | **1** |

* P Value < 0.001

**Figure 3. Active Twitter users at night**

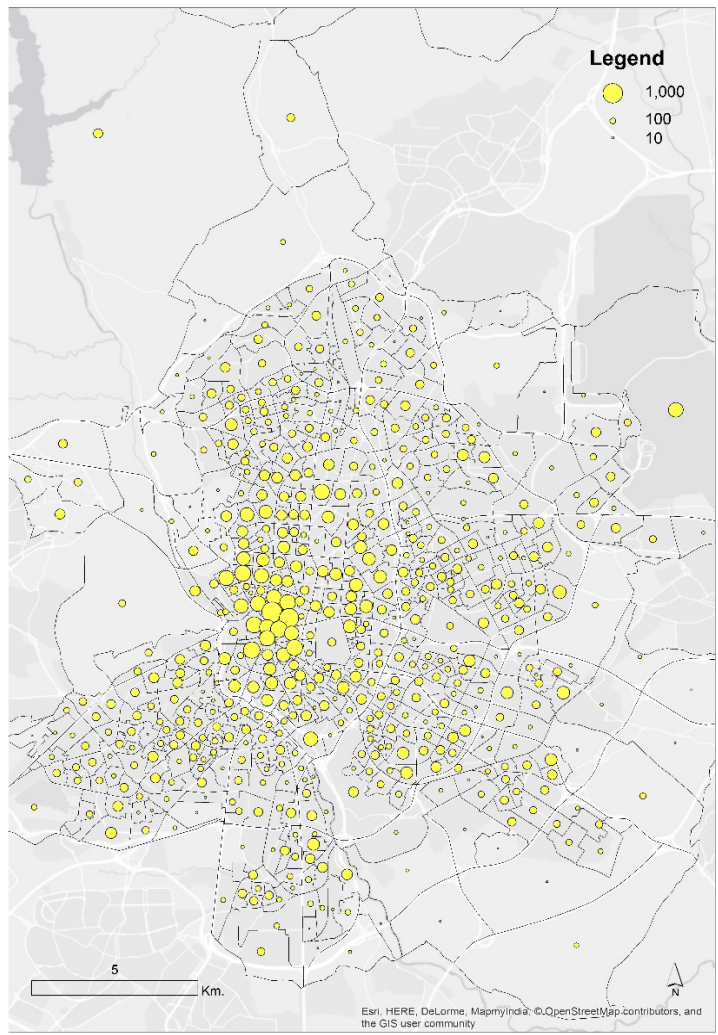

**Figure 4. Residuals in the bivariate correlations of the distribution of active users at night and the rest of the time slots**

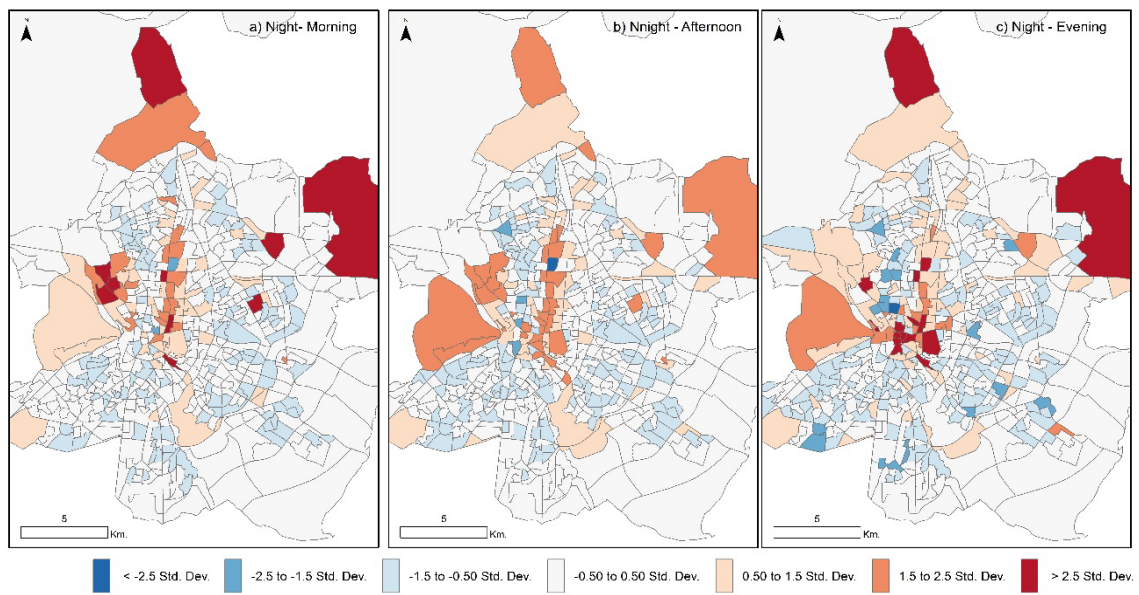

The descriptive statistics show that the lowest standard deviation occurred at night, indicating that active users were distributed throughout all zones in the city (Table 2). The standard deviation increased in the morning, when the population congregated in areas of activity, especially for work, but little use was made of leisure and shopping zones. The highest standard deviation value was recorded in the evening, when the population was more concentrated in the centre, engaged in shopping and leisure activities.

Table 2. Descriptive statistics of the distribution according to time slot

|  | Morning: 08:00 to 13:59 | Afternoon: 14:00 to 18:59 | Evening: 19:00 to 21:59 | Night: 22:00 to 23:59 |
|---|---|---|---|---|
| No. of zones | 584 | 584 | 584 | 584 |
| Minimum: | 1 | 1 | 0 | 0 |
| Maximum | 2015 | 1299 | 1536 | 1099 |
| Total | 100000 | 100000 | 100000 | 100000 |
| Mean: | 171.2 | 171.2 | 171.2 | 171.2 |
| Standard deviation | 166.21 | 150.7 | 169.6 | 141.8 |

**4.2 Temporal Twitter activity profiles according to land use**

Table 3 shows the number of unique users according to land uses and major time slots. Residential land use accounted for the highest number of users. For areas of activity, offices accounted for the highest number of unique users, followed by education and retail. An analysis of data density (normalised unique users / Ha) revealed that the highest values corresponded to land uses that attract a large number of people throughout the day (transportation, culture and retail), whereas parks and industry attracted very low densities.

Table 3. Temporal distribution of active users according to land use and time slot (normalised data)

| Use | Morning: 08:00 to 13:59 | Afternoon: 14:00 to 18:59 | Evening: 19:00 to 21:59 | Night: 22:00 to 23:59 | Total day | Normalised unique user /Ha |
|---|---|---|---|---|---|---|
| Residential | 57022 | 61330 | 61782 | 69784 | 63237 | 4.10 |
| Mixed | 19876 | 19277 | 21061 | 18520 | 18802 | 6.11 |
| Activity total | 23102 | 19393 | 17157 | 11696 | 17961 | 1.28 |
| *Activity:* | | | | | | |
| Retail | 2520 | 2790 | 3046 | 1932 | 2359 | 6.30 |
| Culture | 698 | 644 | 639 | 372 | 623 | 7.16 |
| Education | 4466 | 3187 | 2015 | 1039 | 2653 | 3.44 |
| Industry | 2367 | 1995 | 1730 | 1567 | 2166 | 0.48 |
| Office | 5974 | 4856 | 4289 | 2871 | 4358 | 4.44 |
| Park and sport | 2400 | 2278 | 2403 | 1721 | 2117 | 0.49 |
| Health care | 1378 | 964 | 745 | 473 | 1006 | 4.28 |
| Transport | 1323 | 1059 | 893 | 407 | 1068 | 15.14 |
| Other | 1974 | 1621 | 1398 | 1313 | 1612 | 0.80 |
| Total | 100000 | 100000 | 100000 | 100000 | 100000 | 3.07 |

Table 3 also shows the total number of users by time slot and land use. To facilitate comparisons, the graph in Figure 5 uses relative values to depict very different temporal profiles according to the main types of land use. The curve for residential areas indicates greater activity at night than during the day, while the areas of activity have a much more uneven profile, with a very sharp peak in the morning and a marked drop at night, and as expected, mixed areas show an intermediate situation between the two previous categories.

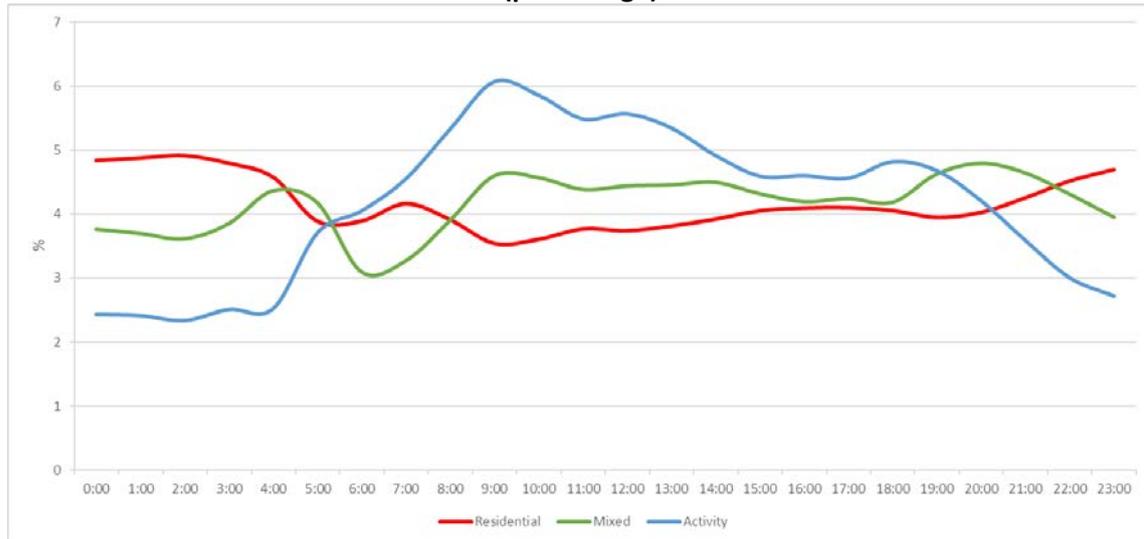

**Figure 5. Temporal distribution of active users according to the main types of land use (percentage)**

The temporal distribution of active users in areas of activity was obtained by combining specific profiles associated with different types of land use (Figure 6). The main transportation infrastructures (airport and railway stations) presented a very early and marked peak. Industry also began very early, although the fluctuations were less pronounced over the course of the day. Education, health and offices showed activity some time later and can be differentiated by the marked peak in the morning in education, which was much less pronounced for offices. Parks and sports areas showed a very stable distribution throughout the day, whereas retail areas were particularly active in the afternoon, after the working day had ended.

**Figure 6. Temporal distribution of active users according to types of activity (percentage)**

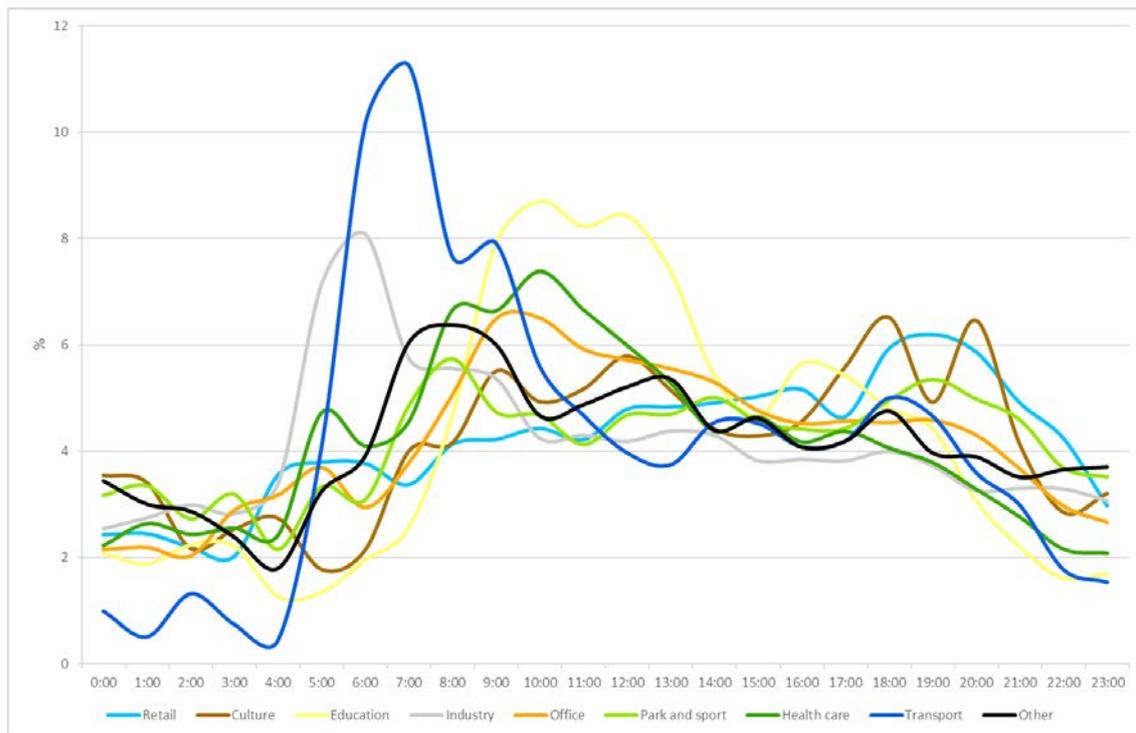

**4.3 Influence of land use on Twitter activity: OLS models**

The results of the OLS models obtained for the different time slots (Table 4) confirmed a close relationship between the spatial distribution of land uses and Twitter activity, since the adjusted coefficients of determination ranged between 0.61 (at night) and 0.76 (in the morning), all with F-statistic values significant at the 0.000 level. Of the explanatory variables considered initially, only two (health and industry) were not significant in all models[2]. At night (from 22 to 24 hours) the variables education, culture and parks were not significant either. Most of the population is at home in this time slot (residential). The presence of categories other than residential in the night model might be due primarily to leisure activities, which were encompassed in the category of others, but were also scattered throughout the city centre in areas where residential use was mixed with other land uses such as offices and retail. Irrespective, the coefficients were very low at night for offices, retail and transportation. The VIF parameter was less than 2 for all significant variables in the four models, indicating that there was no problem of collinearity between the explanatory variables.

As expected, the coefficients of the four models presented positive signs for the various categories of land use and a negative sign for distance to the centre. Thus, the greater the surface area of each land use category, the higher the number of active tweeters, and the further away from the centre, the lower the amount of Twitter activity. The coefficients offer relevant information on the elasticities between number of tweeters and each of the predictor variables. In general, the highest coefficients corresponded to the variables education, culture and retail, indicating that an increase of one unit (one square metre) in these land use categories resulted in a higher increase in tweeters than that recorded for the other categories.

---

[2] Industry is not an activity conducive to workers making use of social networks (in fact, the density of tweets posted from predominantly industrial areas was very low), but this may vary considerably according to the degree to which these spaces are occupied by companies in the service sector (offices). The second case is more surprising, since patients in waiting rooms have time that they could fill by making use of social networks. One explanation for this finding might be the existence of extremely varied health spaces, ranging from large hospitals to private consultants, with a very different intensity of Twitter use.

The variation in the coefficients of the independent variables over the course of the day is consistent with what was observed in the Twitter activity profiles according to land use. The coefficients fell sharply throughout the day in education and transport, whereas this drop was less pronounced in offices and culture. In contrast, the coefficients rose throughout the day in retail, parks and residential zones. The coefficient of the variable distance to the city centre was lower at night, indicating that the centre-periphery gradient of Twitter activity fell at night, given that there is more residential land use on the periphery than in the centre.

Table 4: Results of multiple regression (OLS) models

| Dependent variable: | Coefficients | | | |
|---|---|---|---|---|
| | Morning | Afternoon | Evening | Night |
| Intercept | 123.24* | 112.72* | 125.48* | 110.35* |
| Education (m$^2$) | 0.001909* | 0.001169* | 0.000626* | |
| Office (m$^2$) | 0.000667* | 0.000481* | 0.000443* | 0.000140* |
| Park (m$^2$) | 0.000012* | 0.000016* | 0.000016* | |
| Transport (m$^2$) | 0.000620* | 0.000395* | 0.000347* | 0.000154* |
| Culture (m$^2$) | 0.001640* | 0.001226* | 0.001034* | |
| Retail (m$^2$) | 0.000791* | 0.001044* | 0.001333* | 0.000700* |
| Industry (m$^2$) | | | | |
| Health care (m$^2$) | | | | |
| Residence (m$^2$) | 0.000167* | 0.000228* | 0.000255* | 0.000386* |
| Other (m$^2$) | 0.000055* | 0.000049* | 0.000056* | 0.000044* |
| Distance to city centre (m) | -0.013546* | -0.012342* | -0.015579* | -0.011259* |
| No. Observations | 584 | 584 | 584 | 584 |
| R$^2$ | 0.763* | 0.683* | 0.613* | 0.609* |
| Adj. R$^2$ | 0.759* | 0.678* | 0.609* | 0.605* |
| AIC | 68111 | 6867 | 7121 | 6912 |

* Significant at the 0.01 level
(Variables obtaining non-significant results in the OLS with all variables were discarded and have been left blank)

## 5. FINAL REMARKS

Social networks and big data provide previously unavailable information about urban dynamics, opening up new opportunities in the field of urban studies (Graham and Shelton, 2013). In this study, we used data from one of the most widespread social networks (Twitter) to analyse spatiotemporal demographics in a city. Obviously, the data must be filtered and processed before analysis. In this case, we downloaded and geotagged each of the tweets and aggregated the data spatially and temporally in order to obtain the number of active users according to city zone and time slot. In contrast to other studies that have employed classification algorithms to obtain groups of zones and compare them with land use distribution (inferring land use from Twitter activity), our approach pursues a deeper understanding of the influence of land use on activity in the city. Our groups of land uses were predefined according to a more accurate and useful classification for urban planning than the main groups identified in previous research.

Our analysis of the correlation between the distribution of active tweeters in the night time slot and the different day time slots provides an initial insight into city dynamics, revealing the zones which gain or lose activity in each of the time slots over the course of the day. The former are usually located in the centre and the latter predominate in the periphery, with the logical exception of certain areas of activity such as universities, offices or the airport. Our approach has made it possible to visually identify that offices, education and the main transportation terminals gain activity in the morning, whereas retail areas are most active in the evening.

Subsequently, we linked Twitter activity and land use. First, we performed a descriptive analysis based on typical Twitter activity profiles according to the predominant land use in each transport zone. A marked contrast was observed between residential zones, with increased activity at night, and areas of activity, which were much more active during the day. For most activities, the curve fell from the morning to the evening, except for parks (which had a very stable curve) and retail areas (which showed more activity in the evening than in the morning). Transport terminals and industry presented an early peak which was much more pronounced in the former than in the latter. When the predominant use was education, health or offices, this peak appeared later on in the day, being especially marked in education and less so in offices.

An analysis of zone activity profiles is a simplification, since we only considered the predominant land use in each zone, when most of them host several land uses. Furthermore, this is also a purely descriptive analysis. The OLS analysis examined the influence of the different land uses in each zone on Twitter activity according to time slot. As expected, the coefficients of the independent variables in all models presented positive signs for the land use variables and a negative sign for distance to the centre. The variation in coefficients over the course of the day is consistent with the results obtained for Twitter activity profiles according to predominant land use: rising for residential and retail zones, and falling for education, offices and transport.

Our analyses provide useful information for urban planning, because the results shed light on urban dynamics in relation to land use. This is of interest as regards the provision of services in the public sector (for example, risk assessment and population evacuation plans) and for private sector business activities (potential demand according to zone and time of day). In addition, knowledge of the link between activities and land use can be used to predict future patterns of activity in new urban developments, by using the OLS model to estimate the spatiotemporal distribution of the population according to the envisaged land uses in a new development.

It is clear that social networking data present biases. In the case of Twitter, although it generates an enormous amount of data every day, it is only used by a small percentage of the population. It is also used very unevenly, with some tweeters posting a lot of tweets whereas others only use the network from time to time. This bias can only be partially mitigated, by means of data pre-processing to focus the analysis not on tweets but on the tweeters. Nevertheless, it has been possible to conduct an analysis of the dynamics of city activity and obtain consistent results.